\def\PsfigVersion{1.10}
\def\setDriver{\DvipsDriver} 
\ifx\undefined\psfig\else\endinput\fi
%

\let\LaTeXAtSign=\@
\let\@=\relax
\edef\psfigRestoreAt{\catcode`\@=\number\catcode`@\relax}
\catcode`\@=11\relax
\newwrite\@unused
\def\ps@typeout#1{{\let\protect\string\immediate\write\@unused{#1}}}

\def\DvipsDriver{
	\ps@typeout{psfig/tex \PsfigVersion -dvips}
\def\PsfigSpecials{\DvipsSpecials} 	\def\ps@dir{/}
\def\ps@predir{} }
\def\OzTeXDriver{
	\ps@typeout{psfig/tex \PsfigVersion -oztex}
	\def\PsfigSpecials{\OzTeXSpecials}
	\def\ps@dir{:}
	\def\ps@predir{:}
	\catcode`\^^J=5
}


\def\figurepath{./:}

\def\DoPaths#1{\expandafter\EachPath#1\stoplist}
\def\leer{}
\def\EachPath#1:#2\stoplist{
  \ExistsFile{#1}{\SearchedFile}
  \ifx#2\leer
  \else
    \expandafter\EachPath#2\stoplist
  \fi}
%
%
\def\ps@dir{/}
\def\ExistsFile#1#2{%
   \openin1=\ps@predir#1\ps@dir#2
   \ifeof1
       \closein1
   \else
       \closein1
        \ifx\ps@founddir\leer
           \edef\ps@founddir{#1}
        \fi
   \fi}
%
%
\def\get@dir#1{%
  \def\ps@founddir{}
  \def\SearchedFile{#1}
  \DoPaths\figurepath
}

%
%
\def\@nnil{\@nil}
\def\@empty{}
\def\@psdonoop#1\@@#2#3{}
\def\@psdo#1:=#2\do#3{\edef\@psdotmp{#2}\ifx\@psdotmp\@empty \else
    \expandafter\@psdoloop#2,\@nil,\@nil\@@#1{#3}\fi}
\def\@psdoloop#1,#2,#3\@@#4#5{\def#4{#1}\ifx #4\@nnil \else
       #5\def#4{#2}\ifx #4\@nnil \else#5\@ipsdoloop #3\@@#4{#5}\fi\fi}
\def\@ipsdoloop#1,#2\@@#3#4{\def#3{#1}\ifx #3\@nnil
       \let\@nextwhile=\@psdonoop \else
      #4\relax\let\@nextwhile=\@ipsdoloop\fi\@nextwhile#2\@@#3{#4}}
\def\@tpsdo#1:=#2\do#3{\xdef\@psdotmp{#2}\ifx\@psdotmp\@empty \else
    \@tpsdoloop#2\@nil\@nil\@@#1{#3}\fi}
\def\@tpsdoloop#1#2\@@#3#4{\def#3{#1}\ifx #3\@nnil
       \let\@nextwhile=\@psdonoop \else
      #4\relax\let\@nextwhile=\@tpsdoloop\fi\@nextwhile#2\@@#3{#4}}
%
\ifx\undefined\fbox
\newdimen\fboxrule
\newdimen\fboxsep
\newdimen\ps@tempdima
\newbox\ps@tempboxa
\fboxsep = 3pt
\fboxrule = .4pt
\long\def\fbox#1{\leavevmode\setbox\ps@tempboxa\hbox{#1}\ps@tempdima\fboxrule
    \advance\ps@tempdima \fboxsep \advance\ps@tempdima \dp\ps@tempboxa
   \hbox{\lower \ps@tempdima\hbox
  {\vbox{\hrule height \fboxrule
          \hbox{\vrule width \fboxrule \hskip\fboxsep
          \vbox{\vskip\fboxsep \box\ps@tempboxa\vskip\fboxsep}\hskip
                 \fboxsep\vrule width \fboxrule}
                 \hrule height \fboxrule}}}}
\fi
%
%
\newread\ps@stream
\newif\ifnot@eof       
\newif\if@noisy        
\newif\if@atend        
\newif\if@psfile       
%
%
{\catcode`\%=12\global\gdef\epsf@start{
\def\epsf@PS{PS}
\def\epsf@getbb#1{%
%
%
\openin\ps@stream=\ps@predir#1
\ifeof\ps@stream\ps@typeout{Error, File #1 not found}\else
%
%
   {\not@eoftrue \chardef\other=12
    \def\do##1{\catcode`##1=\other}\dospecials \catcode`\ =10
    \loop
       \if@psfile
	  \read\ps@stream to \epsf@fileline
       \else{
	  \obeyspaces
          \read\ps@stream to \epsf@tmp\global\let\epsf@fileline\epsf@tmp}
       \fi
       \ifeof\ps@stream\not@eoffalse\else
%
%
       \if@psfile\else
       \expandafter\epsf@test\epsf@fileline:. \\%
       \fi
%
%
          \expandafter\epsf@aux\epsf@fileline:. \\%
       \fi
   \ifnot@eof\repeat
   }\closein\ps@stream\fi}%
%
%
\long\def\epsf@test#1#2#3:#4\\{\def\epsf@testit{#1#2}
			\ifx\epsf@testit\epsf@start\else
\ps@typeout{Warning! File does not start with `\epsf@start'.  It may not be a
PostScript file.}
			\fi
			\@psfiletrue} 
%
%
{\catcode`\%=12\global\let\epsf@percent=
%
%
%
\long\def\epsf@aux#1#2:#3\\{\ifx#1\epsf@percent
   \def\epsf@testit{#2}\ifx\epsf@testit\epsf@bblit
	\@atendfalse
        \epsf@atend #3 . \\%
	\if@atend
	   \if@verbose{
		\ps@typeout{psfig: found `(atend)'; continuing search}
	   }\fi
        \else
        \epsf@grab #3 . . . \\%
        \not@eoffalse
        \global\no@bbfalse
        \fi
   \fi\fi}%
%
%
\def\epsf@grab #1 #2 #3 #4 #5\\{%
   \global\def\epsf@llx{#1}\ifx\epsf@llx\empty
      \epsf@grab #2 #3 #4 #5 .\\\else
   \global\def\epsf@lly{#2}%
   \global\def\epsf@urx{#3}\global\def\epsf@ury{#4}\fi}%
%
%
\def\epsf@atendlit{(atend)}
\def\epsf@atend #1 #2 #3\\{%
   \def\epsf@tmp{#1}\ifx\epsf@tmp\empty
      \epsf@atend #2 #3 .\\\else
   \ifx\epsf@tmp\epsf@atendlit\@atendtrue\fi\fi}


\chardef\psletter = 11 
\chardef\other = 12

\newif \ifdebug 
\newif\ifc@mpute 
\c@mputetrue 

\let\then = \relax
\def\r@dian{pt }
\let\r@dians = \r@dian
\let\dimensionless@nit = \r@dian
\let\dimensionless@nits = \dimensionless@nit
\def\internal@nit{sp }
\let\internal@nits = \internal@nit
\newif\ifstillc@nverging
\def \Mess@ge #1{\ifdebug \then \message {#1} \fi}

{ 
	\catcode `\@ = \psletter
	\gdef \nodimen {\expandafter \n@dimen \the \dimen}
	\gdef \term #1 #2 #3%
	       {\edef \t@ {\the #1}
		\edef \t@@ {\expandafter \n@dimen \the #2\r@dian}%
		\t@rm {\t@} {\t@@} {#3}%
	       }
	\gdef \t@rm #1 #2 #3%
	       {{%
		\count 0 = 0
		\dimen 0 = 1 \dimensionless@nit
		\dimen 2 = #2\relax
		\Mess@ge {Calculating term #1 of \nodimen 2}%
		\loop
		\ifnum	\count 0 < #1
		\then	\advance \count 0 by 1
			\Mess@ge {Iteration \the \count 0 \space}%
			\Multiply \dimen 0 by {\dimen 2}%
			\Mess@ge {After multiplication, term = \nodimen 0}%
			\Divide \dimen 0 by {\count 0}%
			\Mess@ge {After division, term = \nodimen 0}%
		\repeat
		\Mess@ge {Final value for term #1 of
				\nodimen 2 \space is \nodimen 0}%
		\xdef \Term {#3 = \nodimen 0 \r@dians}%
		\aftergroup \Term
	       }}
	\catcode `\p = \other
	\catcode `\t = \other
	\gdef \n@dimen #1pt{#1} 
}

\def \Divide #1by #2{\divide #1 by #2} 

\def \Multiply #1by #2
       {{
	\count 0 = #1\relax
	\count 2 = #2\relax
	\count 4 = 65536
	\Mess@ge {Before scaling, count 0 = \the \count 0 \space and
			count 2 = \the \count 2}%
	\ifnum	\count 0 > 32767 
	\then	\divide \count 0 by 4
		\divide \count 4 by 4
	\else	\ifnum	\count 0 < -32767
		\then	\divide \count 0 by 4
			\divide \count 4 by 4
		\else
		\fi
	\fi
	\ifnum	\count 2 > 32767 
	\then	\divide \count 2 by 4
		\divide \count 4 by 4
	\else	\ifnum	\count 2 < -32767
		\then	\divide \count 2 by 4
			\divide \count 4 by 4
		\else
		\fi
	\fi
	\multiply \count 0 by \count 2
	\divide \count 0 by \count 4
	\xdef \product {#1 = \the \count 0 \internal@nits}%
	\aftergroup \product
       }}

\def\r@duce{\ifdim\dimen0 > 90\r@dian \then   
		\multiply\dimen0 by -1
		\advance\dimen0 by 180\r@dian
		\r@duce
	    \else \ifdim\dimen0 < -90\r@dian \then  
		\advance\dimen0 by 360\r@dian
		\r@duce
		\fi
	    \fi}

\def\Sine#1%
       {{%
	\dimen 0 = #1 \r@dian
	\r@duce
	\ifdim\dimen0 = -90\r@dian \then
	   \dimen4 = -1\r@dian
	   \c@mputefalse
	\fi
	\ifdim\dimen0 = 90\r@dian \then
	   \dimen4 = 1\r@dian
	   \c@mputefalse
	\fi
	\ifdim\dimen0 = 0\r@dian \then
	   \dimen4 = 0\r@dian
	   \c@mputefalse
	\fi
	\ifc@mpute \then
		\divide\dimen0 by 180
		\dimen0=3.141592654\dimen0
		\dimen 2 = 3.1415926535897963\r@dian 
		\divide\dimen 2 by 2 
		\Mess@ge {Sin: calculating Sin of \nodimen 0}%
		\count 0 = 1 
		\dimen 2 = 1 \r@dian 
		\dimen 4 = 0 \r@dian 
		\loop
			\ifnum	\dimen 2 = 0 
			\then	\stillc@nvergingfalse
			\else	\stillc@nvergingtrue
			\fi
			\ifstillc@nverging 
			\then	\term {\count 0} {\dimen 0} {\dimen 2}%
				\advance \count 0 by 2
				\count 2 = \count 0
				\divide \count 2 by 2
				\ifodd	\count 2 
				\then	\advance \dimen 4 by \dimen 2
				\else	\advance \dimen 4 by -\dimen 2
				\fi
		\repeat
	\fi
			\xdef \sine {\nodimen 4}%
       }}

\def\Cosine#1{\ifx\sine\UnDefined\edef\Savesine{\relax}\else
		             \edef\Savesine{\sine}\fi
	{\dimen0=#1\r@dian\advance\dimen0 by 90\r@dian
	 \Sine{\nodimen 0}
	 \xdef\cosine{\sine}
	 \xdef\sine{\Savesine}}}

\def\psdraft{
	\def\@psdraft{0}
}
\def\psfull{
	\def\@psdraft{100}
}

\psfull

\newif\if@scalefirst
\def\psscalefirst{\@scalefirsttrue}
\def\psrotatefirst{\@scalefirstfalse}
\psrotatefirst

\newif\if@draftbox
\def\psnodraftbox{
	\@draftboxfalse
}
\def\psdraftbox{
	\@draftboxtrue
}
\@draftboxtrue

\newif\if@prologfile
\newif\if@postlogfile
\def\pssilent{
	\@noisyfalse
}
\def\psnoisy{
	\@noisytrue
}
\psnoisy
\newif\if@bbllx
\newif\if@bblly
\newif\if@bburx
\newif\if@bbury
\newif\if@height
\newif\if@width
\newif\if@rheight
\newif\if@rwidth
\newif\if@angle
\newif\if@clip
\newif\if@verbose
\def\@p@@sclip#1{\@cliptrue}
\newif\if@decmpr
\def\@p@@sfigure#1{\def\@p@sfile{null}\def\@p@sbbfile{null}\@decmprfalse
   \openin1=\ps@predir#1
   \ifeof1
	\closein1
	\get@dir{#1}
	\ifx\ps@founddir\leer
		\openin1=\ps@predir#1.bb
		\ifeof1
			\closein1
			\get@dir{#1.bb}
			\ifx\ps@founddir\leer
				\ps@typeout{Can't find #1 in \figurepath}
			\else
				\@decmprtrue
				\def\@p@sfile{\ps@founddir\ps@dir#1}
				\def\@p@sbbfile{\ps@founddir\ps@dir#1.bb}
			\fi
		\else
			\closein1
			\@decmprtrue
			\def\@p@sfile{#1}
			\def\@p@sbbfile{#1.bb}
		\fi
	\else
		\def\@p@sfile{\ps@founddir\ps@dir#1}
		\def\@p@sbbfile{\ps@founddir\ps@dir#1}
	\fi
   \else
	\closein1
	\def\@p@sfile{#1}
	\def\@p@sbbfile{#1}
   \fi
}
\def\@p@@sfile#1{\@p@@sfigure{#1}}
\def\@p@@sbbllx#1{
		\@bbllxtrue
		\dimen100=#1
		\edef\@p@sbbllx{\number\dimen100}
}
\def\@p@@sbblly#1{
		\@bbllytrue
		\dimen100=#1
		\edef\@p@sbblly{\number\dimen100}
}
\def\@p@@sbburx#1{
		\@bburxtrue
		\dimen100=#1
		\edef\@p@sbburx{\number\dimen100}
}
\def\@p@@sbbury#1{
		\@bburytrue
		\dimen100=#1
		\edef\@p@sbbury{\number\dimen100}
}
\def\@p@@sheight#1{
		\@heighttrue
		\dimen100=#1
   		\edef\@p@sheight{\number\dimen100}
}
\def\@p@@swidth#1{
		\@widthtrue
		\dimen100=#1
		\edef\@p@swidth{\number\dimen100}
}
\def\@p@@srheight#1{
		\@rheighttrue
		\dimen100=#1
		\edef\@p@srheight{\number\dimen100}
}
\def\@p@@srwidth#1{
		\@rwidthtrue
		\dimen100=#1
		\edef\@p@srwidth{\number\dimen100}
}
\def\@p@@sangle#1{
		\@angletrue
		\edef\@p@sangle{#1} 
}
\def\@p@@ssilent#1{
		\@verbosefalse
}
\def\@p@@sprolog#1{\@prologfiletrue\def\@prologfileval{#1}}
\def\@p@@spostlog#1{\@postlogfiletrue\def\@postlogfileval{#1}}
\def\@cs@name#1{\csname #1\endcsname}
\def\@setparms#1=#2,{\@cs@name{@p@@s#1}{#2}}
%
%
\def\ps@init@parms{
		\@bbllxfalse \@bbllyfalse
		\@bburxfalse \@bburyfalse
		\@heightfalse \@widthfalse
		\@rheightfalse \@rwidthfalse
		\def\@p@sbbllx{}\def\@p@sbblly{}
		\def\@p@sbburx{}\def\@p@sbbury{}
		\def\@p@sheight{}\def\@p@swidth{}
		\def\@p@srheight{}\def\@p@srwidth{}
		\def\@p@sangle{0}
		\def\@p@sfile{} \def\@p@sbbfile{}
		\def\@p@scost{10}
		\def\@sc{}
		\@prologfilefalse
		\@postlogfilefalse
		\@clipfalse
		\if@noisy
			\@verbosetrue
		\else
			\@verbosefalse
		\fi
}
%
%
\def\parse@ps@parms#1{
	 	\@psdo\@psfiga:=#1\do
		   {\expandafter\@setparms\@psfiga,}}
%
%
\newif\ifno@bb
\def\bb@missing{
	\if@verbose{
		\ps@typeout{psfig: searching \@p@sbbfile \space  for bounding box}
	}\fi
	\no@bbtrue
	\epsf@getbb{\@p@sbbfile}
        \ifno@bb \else \bb@cull\epsf@llx\epsf@lly\epsf@urx\epsf@ury\fi
}
\def\bb@cull#1#2#3#4{
	\dimen100=#1 bp\edef\@p@sbbllx{\number\dimen100}
	\dimen100=#2 bp\edef\@p@sbblly{\number\dimen100}
	\dimen100=#3 bp\edef\@p@sbburx{\number\dimen100}
	\dimen100=#4 bp\edef\@p@sbbury{\number\dimen100}
	\no@bbfalse
}
\newdimen\p@intvaluex
\newdimen\p@intvaluey
\def\rotate@#1#2{{\dimen0=#1 sp\dimen1=#2 sp
		  \global\p@intvaluex=\cosine\dimen0
		  \dimen3=\sine\dimen1
		  \global\advance\p@intvaluex by -\dimen3
		  \global\p@intvaluey=\sine\dimen0
		  \dimen3=\cosine\dimen1
		  \global\advance\p@intvaluey by \dimen3
		  }}
\def\compute@bb{
		\no@bbfalse
		\if@bbllx \else \no@bbtrue \fi
		\if@bblly \else \no@bbtrue \fi
		\if@bburx \else \no@bbtrue \fi
		\if@bbury \else \no@bbtrue \fi
		\ifno@bb \bb@missing \fi
		\ifno@bb \ps@typeout{FATAL ERROR: no bb supplied or found}
			\no-bb-error
		\fi
		%
%
		\count203=\@p@sbburx
		\count204=\@p@sbbury
		\advance\count203 by -\@p@sbbllx
		\advance\count204 by -\@p@sbblly
		\edef\ps@bbw{\number\count203}
		\edef\ps@bbh{\number\count204}
		\if@angle
			\Sine{\@p@sangle}\Cosine{\@p@sangle}
	        	{\dimen100=\maxdimen\xdef\r@p@sbbllx{\number\dimen100}
					    \xdef\r@p@sbblly{\number\dimen100}
			                    \xdef\r@p@sbburx{-\number\dimen100}
					    \xdef\r@p@sbbury{-\number\dimen100}}
%
                        \def\minmaxtest{
			   \ifnum\number\p@intvaluex<\r@p@sbbllx
			      \xdef\r@p@sbbllx{\number\p@intvaluex}\fi
			   \ifnum\number\p@intvaluex>\r@p@sbburx
			      \xdef\r@p@sbburx{\number\p@intvaluex}\fi
			   \ifnum\number\p@intvaluey<\r@p@sbblly
			      \xdef\r@p@sbblly{\number\p@intvaluey}\fi
			   \ifnum\number\p@intvaluey>\r@p@sbbury
			      \xdef\r@p@sbbury{\number\p@intvaluey}\fi
			   }
			\rotate@{\@p@sbbllx}{\@p@sbblly}
			\minmaxtest
			\rotate@{\@p@sbbllx}{\@p@sbbury}
			\minmaxtest
			\rotate@{\@p@sbburx}{\@p@sbblly}
			\minmaxtest
			\rotate@{\@p@sbburx}{\@p@sbbury}
			\minmaxtest
			\edef\@p@sbbllx{\r@p@sbbllx}\edef\@p@sbblly{\r@p@sbblly}
			\edef\@p@sbburx{\r@p@sbburx}\edef\@p@sbbury{\r@p@sbbury}
		\fi
		\count203=\@p@sbburx
		\count204=\@p@sbbury
		\advance\count203 by -\@p@sbbllx
		\advance\count204 by -\@p@sbblly
		\edef\@bbw{\number\count203}
		\edef\@bbh{\number\count204}
}
%
%
\def\in@hundreds#1#2#3{\count240=#2 \count241=#3
		     \count100=\count240	
		     \divide\count100 by \count241
		     \count101=\count100
		     \multiply\count101 by \count241
		     \advance\count240 by -\count101
		     \multiply\count240 by 10
		     \count101=\count240	
		     \divide\count101 by \count241
		     \count102=\count101
		     \multiply\count102 by \count241
		     \advance\count240 by -\count102
		     \multiply\count240 by 10
		     \count102=\count240	
		     \divide\count102 by \count241
		     \count200=#1\count205=0
		     \count201=\count200
			\multiply\count201 by \count100
		 	\advance\count205 by \count201
		     \count201=\count200
			\divide\count201 by 10
			\multiply\count201 by \count101
			\advance\count205 by \count201
		     \count201=\count200
			\divide\count201 by 100
			\multiply\count201 by \count102
			\advance\count205 by \count201
		     \edef\@result{\number\count205}
}
\def\compute@wfromh{
		\in@hundreds{\@p@sheight}{\@bbw}{\@bbh}
		\edef\@p@swidth{\@result}
}
\def\compute@hfromw{
	        \in@hundreds{\@p@swidth}{\@bbh}{\@bbw}
		\edef\@p@sheight{\@result}
}
\def\compute@handw{
		\if@height
			\if@width
			\else
				\compute@wfromh
			\fi
		\else
			\if@width
				\compute@hfromw
			\else
				\edef\@p@sheight{\@bbh}
				\edef\@p@swidth{\@bbw}
			\fi
		\fi
}
\def\compute@resv{
		\if@rheight \else \edef\@p@srheight{\@p@sheight} \fi
		\if@rwidth \else \edef\@p@srwidth{\@p@swidth} \fi
}
%
\def\compute@sizes{
	\compute@bb
	\if@scalefirst\if@angle
	\if@width
	   \in@hundreds{\@p@swidth}{\@bbw}{\ps@bbw}
	   \edef\@p@swidth{\@result}
	\fi
	\if@height
	   \in@hundreds{\@p@sheight}{\@bbh}{\ps@bbh}
	   \edef\@p@sheight{\@result}
	\fi
	\fi\fi
	\compute@handw
	\compute@resv}
\def\OzTeXSpecials{
	\special{empty.ps /@isp {true} def}
	\special{empty.ps \@p@swidth \space \@p@sheight \space
			\@p@sbbllx \space \@p@sbblly \space
			\@p@sbburx \space \@p@sbbury \space
			startTexFig \space }
	\if@clip{
		\if@verbose{
			\ps@typeout{(clip)}
		}\fi
		\special{empty.ps doclip \space }
	}\fi
	\if@angle{
		\if@verbose{
			\ps@typeout{(rotate)}
		}\fi
		\special {empty.ps \@p@sangle \space rotate \space}
	}\fi
	\if@prologfile
	    \special{\@prologfileval \space } \fi
	\if@decmpr{
		\if@verbose{
			\ps@typeout{psfig: Compression not available
			in OzTeX version \space }
		}\fi
	}\else{
		\if@verbose{
			\ps@typeout{psfig: including \@p@sfile \space }
		}\fi
		\special{epsf=\@p@sfile \space }
	}\fi
	\if@postlogfile
	    \special{\@postlogfileval \space } \fi
	\special{empty.ps /@isp {false} def}
}
\def\DvipsSpecials{
	\special{ps::[begin] 	\@p@swidth \space \@p@sheight \space
			\@p@sbbllx \space \@p@sbblly \space
			\@p@sbburx \space \@p@sbbury \space
			startTexFig \space }
	\if@clip{
		\if@verbose{
			\ps@typeout{(clip)}
		}\fi
		\special{ps:: doclip \space }
	}\fi
	\if@angle
		\if@verbose{
			\ps@typeout{(clip)}
		}\fi
		\special {ps:: \@p@sangle \space rotate \space}
	\fi
	\if@prologfile
	    \special{ps: plotfile \@prologfileval \space } \fi
	\if@decmpr{
		\if@verbose{
			\ps@typeout{psfig: including \@p@sfile.Z \space }
		}\fi
		\special{ps: plotfile "`zcat \@p@sfile.Z" \space }
	}\else{
		\if@verbose{
			\ps@typeout{psfig: including \@p@sfile \space }
		}\fi
		\special{ps: plotfile \@p@sfile \space }
	}\fi
	\if@postlogfile
	    \special{ps: plotfile \@postlogfileval \space } \fi
	\special{ps::[end] endTexFig \space }
}
%
%
\def\psfig#1{\vbox {
	%
	\ps@init@parms
	\parse@ps@parms{#1}
	\compute@sizes
	\ifnum\@p@scost<\@psdraft{
		\PsfigSpecials
		\vbox to \@p@srheight sp{
			\hbox to \@p@srwidth sp{
				\hss
			}
		\vss
		}
	}\else{
		\if@draftbox{
			\hbox{\fbox{\vbox to \@p@srheight sp{
			\vss
			\hbox to \@p@srwidth sp{ \hss
			 \hss }
			\vss
			}}}
		}\else{
			\vbox to \@p@srheight sp{
			\vss
			\hbox to \@p@srwidth sp{\hss}
			\vss
			}
		}\fi

	}\fi
}}
\psfigRestoreAt
\setDriver
\let\@=\LaTeXAtSign


%
%
\def\unredoffs{} \def\redoffs{\voffset=-.31truein\hoffset=-.59truein}
\def\speclscape{}
%
%
%
%
\newbox\leftpage \newdimen\fullhsize \newdimen\hstitle \newdimen\hsbody
\tolerance=1000\hfuzz=2pt\def\fontflag{cm}
\catcode`\@=11 
\def\bigans{b }
\def\answ{b }
%

\ifx\answ\bigans\message{(This will come out unreduced.}
\magnification=1200\unredoffs\baselineskip=16pt plus 2pt minus 1pt
\hsbody=\hsize \hstitle=\hsize 
\else\message{(This will be reduced.} \let\l@r=L
\magnification=1000\baselineskip=16pt plus 2pt minus 1pt \vsize=7truein
\redoffs \hstitle=8truein\hsbody=4.75truein\fullhsize=10truein\hsize=\hsbody
\output={\ifnum\pageno=0 
  \shipout\vbox{\speclscape{\hsize\fullhsize\makeheadline}
   \hbox to \fullhsize{\hfill\pagebody\hfill}}\advancepageno
  \else
 \almostshipout{\leftline{\vbox{\pagebody\makefootline}}}\advancepageno
  \fi}
\def\almostshipout#1{\if L\l@r \count1=1 \message{[\the\count0.\the\count1]}
      \global\setbox\leftpage=#1 \global\let\l@r=R
 \else \count1=2
  \shipout\vbox{\speclscape{\hsize\fullhsize\makeheadline}
      \hbox to\fullhsize{\box\leftpage\hfil#1}}  \global\let\l@r=L\fi}
\fi
%
\newcount\yearltd\yearltd=\year\advance\yearltd by -1900
\def\HUTP#1#2{\Title{HUTP-\number\yearltd/A#1}{#2}}
\def\Title#1#2{\nopagenumbers\abstractfont\hsize=\hstitle\rightline{#1}%
\vskip 1in\centerline{\titlefont #2}\abstractfont\vskip .5in\pageno=0}
\def\Date#1{\vfill\leftline{#1}\tenpoint\supereject\global\hsize=\hsbody%
\footline={\hss\tenrm\folio\hss}}
%
\def\draft{\draftmode\Date{\draftdate}}
\def\draftmode{\message{ DRAFTMODE }\def\draftdate{{\rm preliminary draft:
\number\month/\number\day/\number\yearltd\ \ \hourmin}}%
\headline={\hfil\draftdate}\writelabels\baselineskip=20pt plus 2pt minus 2pt
 {\count255=\time\divide\count255 by 60 \xdef\hourmin{\number\count255}
  \multiply\count255 by-60\advance\count255 by\time
  \xdef\hourmin{\hourmin:\ifnum\count255<10 0\fi\the\count255}}}
\def\nolabels{\def\wrlabeL##1{}\def\eqlabeL##1{}\def\reflabeL##1{}}
\def\writelabels{\def\wrlabeL##1{\leavevmode\vadjust{\rlap{\smash%
{\line{{\escapechar=` \hfill\rlap{\sevenrm\hskip.03in\string##1}}}}}}}%
\def\eqlabeL##1{{\escapechar-1\rlap{\sevenrm\hskip.05in\string##1}}}%
\def\reflabeL##1{\noexpand\llap{\noexpand\sevenrm\string\string\string##1}}}
\nolabels
%
\global\newcount\secno \global\secno=0
\global\newcount\meqno \global\meqno=1
\def\newsec#1{\global\advance\secno by1\message{(\the\secno. #1)}
\global\subsecno=0\eqnres@t\noindent{\bf\the\secno. #1}
\writetoca{{\secsym} {#1}}\par\nobreak\medskip\nobreak}
\def\eqnres@t{\xdef\secsym{\the\secno.}\global\meqno=1\bigbreak\bigskip}
\def\sequentialequations{\def\eqnres@t{\bigbreak}}\xdef\secsym{}
\global\newcount\subsecno \global\subsecno=0
\def\subsec#1{\global\advance\subsecno by1\message{(\secsym\the\subsecno. #1)}
\ifnum\lastpenalty>9000\else\bigbreak\fi
\noindent{\it\secsym\the\subsecno. #1}\writetoca{\string\quad
{\secsym\the\subsecno.} {#1}}\par\nobreak\medskip\nobreak}
\def\appendix#1#2{\global\meqno=1\global\subsecno=0\xdef\secsym{\hbox{#1.}}
\bigbreak\bigskip\noindent{\bf Appendix #1. #2}\message{(#1. #2)}
\writetoca{Appendix {#1.} {#2}}\par\nobreak\medskip\nobreak}
%
%
\def\eqnn#1{\xdef #1{(\secsym\the\meqno)}\writedef{#1\leftbracket#1}%
\global\advance\meqno by1\wrlabeL#1}
\def\eqna#1{\xdef #1##1{\hbox{$(\secsym\the\meqno##1)$}}
\writedef{#1\numbersign1\leftbracket#1{\numbersign1}}%
\global\advance\meqno by1\wrlabeL{#1$\{\}$}}
\def\eqn#1#2{\xdef #1{(\secsym\the\meqno)}\writedef{#1\leftbracket#1}%
\global\advance\meqno by1$$#2\eqno#1\eqlabeL#1$$}
%
\newskip\footskip\footskip14pt plus 1pt minus 1pt 
\def\footnotefont{\ninepoint}\def\f@t#1{\footnotefont #1\@foot}
\def\f@@t{\baselineskip\footskip\bgroup\footnotefont\aftergroup\@foot\let\next}
\setbox\strutbox=\hbox{\vrule height9.5pt depth4.5pt width0pt}
\global\newcount\ftno \global\ftno=0
\def\foot{\global\advance\ftno by1\footnote{$^{\the\ftno}$}}
%
\newwrite\ftfile
\def\footend{\def\foot{\global\advance\ftno by1\chardef\wfile=\ftfile
$^{\the\ftno}$\ifnum\ftno=1\immediate\openout\ftfile=foots.tmp\fi%
\immediate\write\ftfile{\noexpand\smallskip%
\noexpand\item{f\the\ftno:\ }\pctsign}\findarg}%
\def\footatend{\vfill\eject\immediate\closeout\ftfile{\parindent=20pt
\centerline{\bf Footnotes}\nobreak\bigskip\input foots.tmp }}}
\def\footatend{}
%
%
\global\newcount\refno \global\refno=1
\newwrite\rfile
\def\ref{[\the\refno]\nref}
\def\nref#1{\xdef#1{[\the\refno]}\writedef{#1\leftbracket#1}%
\ifnum\refno=1\immediate\openout\rfile=refs.tmp\fi
\global\advance\refno by1\chardef\wfile=\rfile\immediate
\write\rfile{\noexpand\item{#1\ }\reflabeL{#1\hskip.31in}\pctsign}\findarg}
\def\findarg#1#{\begingroup\obeylines\newlinechar=`\^^M\pass@rg}
{\obeylines\gdef\pass@rg#1{\writ@line\relax #1^^M\hbox{}^^M}%
\gdef\writ@line#1^^M{\expandafter\toks0\expandafter{\striprel@x #1}%
\edef\next{\the\toks0}\ifx\next\em@rk\let\next=\endgroup\else\ifx\next\empty%
\else\immediate\write\wfile{\the\toks0}\fi\let\next=\writ@line\fi\next\relax}}
\def\striprel@x#1{} \def\em@rk{\hbox{}}
\def\lref{\begingroup\obeylines\lr@f}
\def\lr@f#1#2{\gdef#1{\ref#1{#2}}\endgroup\unskip}
\def\semi{;\hfil\break}
\def\addref#1{\immediate\write\rfile{\noexpand\item{}#1}} 
\def\listrefs{\footatend\vfill\supereject\immediate\closeout\rfile\writestoppt
\baselineskip=14pt\centerline{{\bf References}}\bigskip{\frenchspacing%
\parindent=20pt\escapechar=` \input refs.tmp\vfill\eject}\nonfrenchspacing}
\def\startrefs#1{\immediate\openout\rfile=refs.tmp\refno=#1}
\def\xref{\expandafter\xr@f}\def\xr@f[#1]{#1}
\def\refs#1{\count255=1[\r@fs #1{\hbox{}}]}
\def\r@fs#1{\ifx\und@fined#1\message{reflabel \string#1 is undefined.}%
\nref#1{need to supply reference \string#1.}\fi%
\vphantom{\hphantom{#1}}\edef\next{#1}\ifx\next\em@rk\def\next{}%
\else\ifx\next#1\ifodd\count255\relax\xref#1\count255=0\fi%
\else#1\count255=1\fi\let\next=\r@fs\fi\next}
\def\figures{\centerline{{\bf Figure Captions}}\medskip\parindent=40pt%
\def\fig##1##2{\medskip\item{Figure~##1.  }##2}}
%
\newwrite\ffile\global\newcount\figno \global\figno=1
\def\fig{Figure~\the\figno\nfig}
\def\nfig#1{\xdef#1{Figure~\the\figno}%
\writedef{#1\leftbracket fig.\noexpand~\the\figno}%
\ifnum\figno=1\immediate\openout\ffile=figs.tmp\fi\chardef\wfile=\ffile%
\immediate\write\ffile{\noexpand\medskip\noexpand\item{Fig.\ \the\figno. }
\reflabeL{#1\hskip.55in}\pctsign}\global\advance\figno by1\findarg}
\def\listfigs{\vfill\eject\immediate\closeout\ffile{\parindent40pt
\baselineskip14pt\centerline{{\bf Figure Captions}}\nobreak\medskip
\escapechar=` \input figs.tmp\vfill\eject}}
\def\xfig{\expandafter\xf@g}\def\xf@g fig.\penalty\@M\ {}
\def\figs#1{figs.~\f@gs #1{\hbox{}}}
\def\f@gs#1{\edef\next{#1}\ifx\next\em@rk\def\next{}\else
\ifx\next#1\xfig #1\else#1\fi\let\next=\f@gs\fi\next}
\newwrite\lfile
{\escapechar-1\xdef\pctsign{\string\%}\xdef\leftbracket{\string\{}
\xdef\rightbracket{\string\}}\xdef\numbersign{\string\#}}
\def\writedefs{\immediate\openout\lfile=labeldefs.tmp \def\writedef##1{%
\immediate\write\lfile{\string\def\string##1\rightbracket}}}
\def\writestop{\def\writestoppt{\immediate\write\lfile{\string\pageno%
\the\pageno\string\startrefs\leftbracket\the\refno\rightbracket%
\string\def\string\secsym\leftbracket\secsym\rightbracket%
\string\secno\the\secno\string\meqno\the\meqno}\immediate\closeout\lfile}}
\def\writestoppt{}\def\writedef#1{}
\def\seclab#1{\xdef #1{\the\secno}\writedef{#1\leftbracket#1}\wrlabeL{#1=#1}}
\def\subseclab#1{\xdef #1{\secsym\the\subsecno}%
\writedef{#1\leftbracket#1}\wrlabeL{#1=#1}}
\newwrite\tfile \def\writetoca#1{}
\def\leaderfill{\leaders\hbox to 1em{\hss.\hss}\hfill}
\def\writetoc{\immediate\openout\tfile=toc.tmp
   \def\writetoca##1{{\edef\next{\write\tfile{\noindent ##1
   \string\leaderfill {\noexpand\number\pageno} \par}}\next}}}
\def\listtoc{\centerline{\bf Contents}\nobreak\medskip{\baselineskip=12pt
 \parskip=0pt\catcode`\@=11 \input toc.tex \catcode`\@=12 \bigbreak\bigskip}}
\catcode`\@=12 
%
\edef\tfontsize{\ifx\answ\bigans scaled\magstep3\else scaled\magstep4\fi}
\font\titlerm=cmr10 \tfontsize \font\titlerms=cmr7 \tfontsize
\font\titlermss=cmr5 \tfontsize \font\titlei=cmmi10 \tfontsize
\font\titleis=cmmi7 \tfontsize \font\titleiss=cmmi5 \tfontsize
\font\titlesy=cmsy10 \tfontsize \font\titlesys=cmsy7 \tfontsize
\font\titlesyss=cmsy5 \tfontsize \font\titleit=cmti10 \tfontsize
\skewchar\titlei='177 \skewchar\titleis='177 \skewchar\titleiss='177
\skewchar\titlesy='60 \skewchar\titlesys='60 \skewchar\titlesyss='60
\def\titlefont{\def\rm{\fam0\titlerm}
\textfont0=\titlerm \scriptfont0=\titlerms \scriptscriptfont0=\titlermss
\textfont1=\titlei \scriptfont1=\titleis \scriptscriptfont1=\titleiss
\textfont2=\titlesy \scriptfont2=\titlesys \scriptscriptfont2=\titlesyss
\textfont\itfam=\titleit \def\it{\fam\itfam\titleit}\rm}
\font\authorfont=cmcsc10 \ifx\answ\bigans\else scaled\magstep1\fi
\ifx\answ\bigans\def\abstractfont{\tenpoint}\else
\font\abssl=cmsl10 scaled \magstep1
\font\absrm=cmr10 scaled\magstep1 \font\absrms=cmr7 scaled\magstep1
\font\absrmss=cmr5 scaled\magstep1 \font\absi=cmmi10 scaled\magstep1
\font\absis=cmmi7 scaled\magstep1 \font\absiss=cmmi5 scaled\magstep1
\font\abssy=cmsy10 scaled\magstep1 \font\abssys=cmsy7 scaled\magstep1
\font\abssyss=cmsy5 scaled\magstep1 \font\absbf=cmbx10 scaled\magstep1
\skewchar\absi='177 \skewchar\absis='177 \skewchar\absiss='177
\skewchar\abssy='60 \skewchar\abssys='60 \skewchar\abssyss='60
\def\abstractfont{\def\rm{\fam0\absrm}
\textfont0=\absrm \scriptfont0=\absrms \scriptscriptfont0=\absrmss
\textfont1=\absi \scriptfont1=\absis \scriptscriptfont1=\absiss
\textfont2=\abssy \scriptfont2=\abssys \scriptscriptfont2=\abssyss
\textfont\itfam=\bigit \def\it{\fam\itfam\bigit}\def\footnotefont{\tenpoint}%
\textfont\slfam=\abssl \def\sl{\fam\slfam\abssl}%
\textfont\bffam=\absbf \def\bf{\fam\bffam\absbf}\rm}\fi
\def\tenpoint{\def\rm{\fam0\tenrm}
\textfont0=\tenrm \scriptfont0=\sevenrm \scriptscriptfont0=\fiverm
\textfont1=\teni  \scriptfont1=\seveni  \scriptscriptfont1=\fivei
\textfont2=\tensy \scriptfont2=\sevensy \scriptscriptfont2=\fivesy
\textfont\itfam=\tenit \def\it{\fam\itfam\tenit}\def\footnotefont{\ninepoint}%
\textfont\bffam=\tenbf \def\bf{\fam\bffam\tenbf}\def\sl{\fam\slfam\tensl}\rm}
\font\ninerm=cmr9 \font\sixrm=cmr6 \font\ninei=cmmi9 \font\sixi=cmmi6
\font\ninesy=cmsy9 \font\sixsy=cmsy6 \font\ninebf=cmbx9
\font\nineit=cmti9 \font\ninesl=cmsl9 \skewchar\ninei='177
\skewchar\sixi='177 \skewchar\ninesy='60 \skewchar\sixsy='60
\def\ninepoint{\def\rm{\fam0\ninerm}
\textfont0=\ninerm \scriptfont0=\sixrm \scriptscriptfont0=\fiverm
\textfont1=\ninei \scriptfont1=\sixi \scriptscriptfont1=\fivei
\textfont2=\ninesy \scriptfont2=\sixsy \scriptscriptfont2=\fivesy
\textfont\itfam=\ninei \def\it{\fam\itfam\nineit}\def\sl{\fam\slfam\ninesl}%
\textfont\bffam=\ninebf \def\bf{\fam\bffam\ninebf}\rm}
%
%
\def\noblackbox{\overfullrule=0pt}
\hyphenation{anom-aly anom-alies coun-ter-term coun-ter-terms}
\def\inv{^{\raise.15ex\hbox{${\scriptscriptstyle -}$}\kern-.05em 1}}
\def\dup{^{\vphantom{1}}}
\def\Dsl{\,\raise.15ex\hbox{/}\mkern-13.5mu D} 
\def\dsl{\raise.15ex\hbox{/}\kern-.57em\partial}
\def\del{\partial}
\def\Psl{\dsl}
\def\tr{{\rm tr}} \def\Tr{{\rm Tr}}
\font\bigit=cmti10 scaled \magstep1
\def\biglie{\hbox{\bigit\$}} 
\def\lspace{\ifx\answ\bigans{}\else\qquad\fi}
\def\lbspace{\ifx\answ\bigans{}\else\hskip-.2in\fi} 
\def\boxeqn#1{\vcenter{\vbox{\hrule\hbox{\vrule\kern3pt\vbox{\kern3pt
	\hbox{${\displaystyle #1}$}\kern3pt}\kern3pt\vrule}\hrule}}}
\def\mbox#1#2{\vcenter{\hrule \hbox{\vrule height#2in
		\kern#1in \vrule} \hrule}}  
%
\def\CAG{{\cal A/\cal G}} \def\CO{{\cal O}} 
\def\CA{{\cal A}} \def\CC{{\cal C}} \def\CF{{\cal F}} \def\CG{{\cal G}}
\def\CL{{\cal L}} \def\CH{{\cal H}} \def\CI{{\cal I}} \def\CU{{\cal U}}
\def\CB{{\cal B}} \def\CR{{\cal R}} \def\CD{{\cal D}} \def\CT{{\cal T}}
\def\e#1{{\rm e}^{^{\textstyle#1}}}
\def\grad#1{\,\nabla\!_{{#1}}\,}
\def\gradgrad#1#2{\,\nabla\!_{{#1}}\nabla\!_{{#2}}\,}
\def\ph{\varphi}
\def\psibar{\overline\psi}
\def\om#1#2{\omega^{#1}{}_{#2}}
\def\vev#1{\langle #1 \rangle}
\def\lform{\hbox{$\sqcup$}\llap{\hbox{$\sqcap$}}}
\def\darr#1{\raise1.5ex\hbox{$\leftrightarrow$}\mkern-16.5mu #1}
\def\lie{\hbox{\it\$}} 
\def\ha{{1\over2}}
\def\half{{\textstyle{1\over2}}} 
\def\roughly#1{\raise.3ex\hbox{$#1$\kern-.75em\lower1ex\hbox{$\sim$}}}

\def\nv{{\bf n}}
\def\Nv{{\bf N}}
\def\xv{{\bf x}}
\def\cv{{\bf c}}
\def\ev{{\bf e}}
\def\Rv{{\bf R}}
\def\lv{{\bf l}}
\def\av{{\bf a}}
\def\uv{{\bf u}}
\def\gradv{{\bf \nabla}}
\def\kbT{k_{\scriptscriptstyle\rm B}T}
\Title{}{Chiral Fluctuations and Structures}
\pageno=1
\footline={\hss\tenrm\folio\hss}
\centerline{T.C.~Lubensky, Randall D. Kamien and Holger Stark}
\smallskip\centerline{\sl Department of Physics and Astronomy,
University of Pennsylvania,}
\centerline{\sl Philadelphia, PA 19104}
\centerline{22 December 1995}

\vskip .3in
\vbox{\baselineskip=0.16truein
\noindent Chiral molecules form a number of non-chiral structures, the simplest
being
an isotropic fluid phase.  In a mesophase of achiral
molecules the fluctuations will on average be achiral as well: left-handed
twists and right-handed twists will occur with the same probability.
In a system composed of chiral molecules, however, the fluctuations will be
biased
in one direction or the other, and there will be optical effects such
as a non-zero
fluctuation induced
rotary power {\sl in addition} to the molecular rotary power of the
individual molecules.
We discuss this effect in a number of contexts, including
lyotropic lamellar phases.  When the tendency to twist is strong enough
or the line energy of having an exposed edge is small enough, a new, defect
riddled ground state of a lyotropic lamella, akin to the Renn-Lubensky
twist-grain-boundary phase of thermotropic smectics, can occur.
}

\newsec{Introduction}

Since the discovery of rotary power by Biot in the early 19$^{\rm th}$ century,
we have expanded
our understanding of the form and function of chiral molecules
with almost reckless abandon.  The microscopic origin of macroscopic chiral
structures is still not very well understood.
Chiral molecules have no reflection planes as shown in Figure 1.
We also show achiral molecules from which chiral molecules
can be formed by chemical changes.
A fundamental feature of all chiral molecules is that they are
three dimensional:  a planar molecule is invariant with respect to
reflections through the plane in which it lies. The loss of reflection
symmetry
causes energetic and entropic preference for axes of neighboring molecules
to rotate relative to each other.  This leads to a new length scale (the
inverse of the rotation angle per unit distance) normally much larger than
any molecular length and to a spectacular array of equilibrium spatially
modulated chiral phases, including the cholesteric
(or chiral nematic $N^*$)
phase, the blue phases, the TGB phases, and the smectic-$C^*$ phase.
These equilibrium phases, like the molecules of which they are comprised,
have
no reflection planes.

\bigskip
\vbox{\hoffset=-0.5truein
\centerline{\hskip 0.5truein\vbox{\hsize=4truein\baselineskip=0.16truein
\psfig{figure=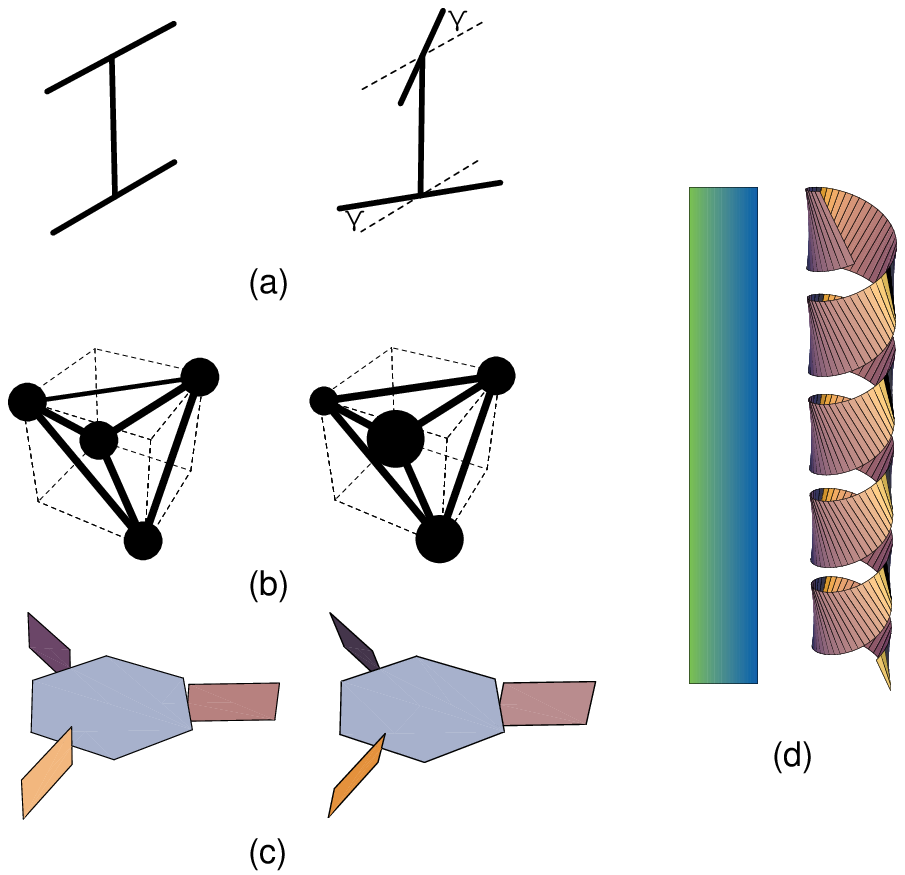}
\noindent {\bf Figure 1.} This figure shows a series of achiral molecules and
the chiral
molecules they become after an appropriate transformation. (a) left, a
planar
achiral $H$-shaped molecule; right a chiral molecule obtained by rotating
one
leg of the $H$ relative to the other through an angle between $0$ and
$\pi/2$.
(b) left: a tetrahedral molecule like $CH_4$ with identical molecules at
all
four vertices; right a chiral molecule with different atoms at each vertex.
(c) left: an achiral discotic molecule with three propeller blades;
right:
a chiral discotic molecule obtained from the molecule at the right rotating
through an angle between $0$ and $\pi/2$
(d)
left: a planar ``strip" molecule; right: a chiral helical ribbon obtained
by
wrapping the strip at the right around a cylinder.  This is a model
for
DNA.
}}}
\bigskip

Chiral molecules can also form phases which do not
exhibit macroscopic chiral order, {\sl i.e.}, phases that do have reflection
planes.
Examples of non-chiral phases composed of chiral molecules are the
isotropic phase and the smectic-$A$ phase. Even though the structure of
these
phases is achiral, they nevertheless exhibit chiral properties if they are
composed of chiral molecules.  In particular, there will be chiral
fluctuations,
the most familiar manifestation of which is optical
rotary power of a
pre-cholesteric
isotropic phase of chiral nematogens \ref\chicho{D.~Bensimon, E.~Domany
and S.~Shtrikman, Phys. Rev. A {\bf 28}, 427 (1982).}.   Of course
the individual molecules have a non-zero rotary power ({\sl e.g.} a solution
of sugar molecules rotates light) as well.  Since the optical chirality
of the molecules need {\sl not} be the same as the chirality of
the resulting mesophase, the chiral fluctuations can lead to either
a reduction or enhancement of the rotary power over that of a very dilute
solution.

This talk will explore possible new structures produced by chirality and
fluctuations in achiral phases of chiral molecules.

\newsec{Chiral Fluctuations in Achiral Phases}

The isotropic and smectic-$A$ phases are achiral liquid
crystal mesophases that can be, nonetheless, composed of
chiral molecules.  Residual chiral fluctuations can lead, however
to optical rotary power.
Other achiral states exist at isolated points of
the phase diagram of smectic-$C^*$ phases and cholesterics. Some of these
mesophases have the
remarkable property of twist inversion
\ref\twistinver{
A.J.~Slaney, I.~Nishiyama, P.~Styring and J.W.~Goodby, J. Mater. Chem.
{\bf 2}, 805 (1992); P.~Styring, J.D.~Vuijk, I.~Nishiyama, A.J.~Slaney
and J.W.~Goodby, J. Mater. Chem. {\bf 3},
399 (1993); I.~Dierking, F.~Gie\ss elmann, P.~Zugenmaier, K.~Mohr, H.~Zaschke
and W.~Kuczynski, Liq. Cryst. {\bf 18}, 443 (1995).},
where the mesoscopic pitch changes sign without a concomitant microscopic
change in the molecular structure.

\subsec{The Smectic-$A$ Phase}

The smectic-$A$ phase, like the isotropic phase, has an achiral structure.
Its
periodic layering expels molecular twist \ref\dg{
P.G.~de~Gennes, Solid State Commun. {\bf 14}, 997 (1973).}.  Thus, chiral
molecules can and
often
do form an achiral smectic-$A$ phase.
Around the achiral ground state there will be chiral fluctuations.
To
study these fluctuations, we use the low-temperature elastic free energy of
a
smectic, expressed in terms of the layer displacement variable $u$ and the
Frank director $\nv = \nv_0 + \delta \nv$ with $\nv_0 = {\bf\hat z}$:
\eqn\efreeone{
F = F_u + F_n + F_{\rm Ch} ,
}
where
\eqn\fu
{F_u = {1\over 2}\int d^3\!x\, [B (\nabla_{||} u )^2 + D( \gradv_{\perp} u +
\delta
\nv )^2 ]
}
is the strain energy,
\eqn\fs{
F_n = {1\over 2}\int d^3\!x\, \left\{ K_1 ( \gradv \cdot \nv )^2 + K_2 [\nv
\cdot
(\gradv \times \nv)]^2 + K_3 [\nv \times ( \gradv \times \nv)]^2 \right\}
}
is the Frank free energy, and
\eqn\fch{
F_{\rm Ch} =  K_2k_0 \int d^3\!x\, \nv \cdot ( \gradv \times \nv )
}
is the chiral energy.  In the last equation $k_0=2\pi/P$, where $P$ is the
cholesteric
pitch.

{}From this free energy, we can calculate the correlations function of the
displacement and director fields. We find in the special case that $B=D$
and $K=K_1=K_2=K_3$ (with $q^2=q_\perp^2+q_z^2$) \ref\KLS{R.D.~Kamien,
T.C.~Lubensky and H.~Stark, {\sl in preparation} (1996).}
\eqn\SmAfluc{
G_{\delta n_i \delta n_j}(q_\perp,q_z) = \kbT{(B+Kq^2)q^2\delta_{ij}
- iKk_0q_zq^2\epsilon_{ij}\over
(B+Kq^2)^2q^2 - K^2k_0^2q_z^2q^2-B(B+Kq^2)q_\perp^2}
}
The terms linear in $k_0$ represent chiral fluctuations since they
multiply the two-dimensional antisymmetric tensor.  It should be
possible to measure these quantities via light scattering and to distinguish
between a smectic composed of chiral molecules and one that is not.

The chiral
director fluctuations are nearly frozen out: they are only correlated
on the scale of the twist penetration depth $\lambda=\sqrt{K/B}$.
Nevertheless,
these short range correlations
lead to a fluctuation induced rotary power, which for light travelling
along the $z$ direction is
\eqn\SmArotpower{R= {\lambda k_0}\left[\omega\over c\right]^2 {\kbT\over 48\pi
K}
{\left(\epsilon_\perp-
\epsilon_{\scriptscriptstyle ||}\right)^2\over
\epsilon_\perp
} {\hbox{\sl radians}\over\hbox{\sl cm}}}
where $\omega$ is the frequency of the polarized light beam, $c$ is the speed
of light in vacuum and $\epsilon_\perp$ and $\epsilon_{\scriptscriptstyle ||}$
are
the perpendicular and parallel dielectric constants of the mesogens.  Putting
in
typical numbers
we get rotations between $5\times 10^{-2}$ and $0.5$
{\sl radians/cm} deep in the smectic phase.  As
the smectic phase melts and forms a cholesteric, $B\rightarrow 0$ and thus the
rotary power will grow
as $1/\sqrt{B}$.  At the transition the fluctuation induced rotary power
will grow larger than the molecular rotary power (roughly $10^{-1}${\sl
radians/cm}).
This effect could also be quite large in colloidal liquid crystals which
have penetration depths on the order of microns.

\subsec{Compensated Cholesterics}

There are a number of molecules
\ref\compchol{H.~Stegemeyer, K.~Siemensmeyer, W.~Sucrow and L.~Appel, Z.
Naturforsch.
{\bf 44a}, 1127 (1989).} whose cholesteric phases
exhibit the phenomenon of twist inversion. Their pitch wavenumber $k_0$
passes through zero and changes sign as a function of temperature or other
external parameter.  At the point where $k_0=0$, the pitch is $\infty$, and
the resultant ``cholesteric" phase has the same macroscopic symmetry as an
achiral nematic.  The molecules, however, are still chiral, and this
nematic
phase will exhibit chiral fluctuations not present in the achiral nematic
phase.  The microscopic origin of twist inversion is not totally
understood.  Any microscopic mechanism will, however, eventually lead to a
phenomenological free energy for the director $\nv$ and some other variable
whose variation can lead to a vanishing of $k_0$.  A simple model for a
cholesteric that displays twist inversion can be constructed from $\nv$ and
$S$, the Maier-Saupe order parameter, which measures
the magnitude of the nematic order parameter:
\eqn\efreetwo
{F= F_n + \int d^3\!x\, f(S) - \int d^3\!x\, ( a S^2 - b S^4 ) \nv \cdot
(\gradv
\times \nv ) ,
}
where $f(S)$ is a free energy density depending only on $S$.
Compensation occurs when the coefficient of $\nv \cdot ( \gradv \times \nv
)$ is zero, i.e., when $S^2 = a/b$.  Near this point, we can write $S =
(a/b)^{1/2} + \delta S$ and
\eqn\efreethree{
F = F_n + F_0 + {1\over 2} \int d^3\!x\, A (\delta S)^2 - \alpha\int d^3 \!x\,
\delta S \nv \cdot (\gradv \times \nv ) ,
}
where $A= f^{\prime\prime}(\sqrt{a/b})$ and $\alpha = 2 a^2 /b$.

While there will be, on average, no net chiral fluctuations in the director,
there is still a fundamental asymmetry.  Fluctuations with one handedness
(or one sign of $\nv\cdot\left(\nabla\!\times\!\nv\right)$) are coupled
to a decrease in $S$ while fluctuations of the other handedness go with
an increase of $S$.
Since fluctuations in both $S$ and $n$ contribute to fluctuations in the
dielectric tensor, the chiral terms in the above correlation functions
will lead to a nonzero fluctuation-induced rotary power of a compensated
cholesteric.

\subsec{Compensated Smectic-$C^*$ Phases}

The chiral smectic-$C^*$ phase, like the cholesteric phase can exhibit
twist
inversion.  A simple model free energy producing this phenomenon can be
constructed from the magnitude and phase of the $\cv$-director:
$\cv = c(\cos\theta , \sin\theta , 0 )$:
\eqn\SmCchiral{
F = \int d^3\!x\, {1\over 2} K (\gradv \theta )^2 - \int d^3\!x\, (a c^2 - b
c^4)
\nabla_z \theta + \int d^3\!x\, f( c ) ,}
where $f(c)$ is a free energy density.
The chiral wavenumber is zero when $c^2 = a/b$. Expanding about this point,
we
obtain
\eqn\efreex{
F = {1\over 2} \int d^3\!x\,\left\{A(\delta c )^2 + K ( \gradv \theta )^2 -
2\sqrt{a^3/b}\delta c \nabla_z \theta \right\} ,
}
This leads to
\eqn\ecorrfour{
G_{ab} = {1 \over A K q^2 - (a^3/b)q_z^2}
\pmatrix{K q^2 & i \sqrt{a^3/b} q_z\cr
         -i\sqrt{a^3/b} q_z & A \cr} .
}
The rotary power will again scale as
\eqn\erotagain{
R \sim \lambda \sqrt{a^3\over b}\left[{\omega\over
c}\right]^2{\left(\epsilon_\perp-\epsilon_{\scriptscriptstyle ||}\right)^2\over
2\epsilon_\perp+\epsilon_{\scriptscriptstyle ||}} }

\nref\SL{S.~Langer and J.~Sethna, Phys. Rev. A {\bf 34}
(1986) 5035; G.A.~Hinshaw, Jr., R.G.~Petschek and R.A.~Pelcovits, Phys. Rev.
Let
t.
{\bf 60}
(1988) 1864.}
\nref\SEL{J.V.~Selinger and J.M.~Schnur, Phys. Rev. Lett. {\bf 71}
(1993) 4091; J.V. Selinger, Z.-G. Wang, R.F.~Bruinsma and C.M.~Knobler,
Phys. Rev. Lett. {\bf 70} (1993) 1139.}

Equation \SmCchiral\ is the simplest model free energy for a
compensated
smectic-$C^*$ phase. It can be made more realistic and considerably
more complicated by
including such things as the anisotropy of the Frank elastic constants and
the couplings to the electric dipole moment, which is always
present in chiral
ferroelectric liquid crystals.

The chiral term proportional to
$\nabla_z \theta$ arises from a $\cv \cdot (\gradv \times \cv )$ term in
the
free energy.  There are also other chiral terms such as $\Nv \cdot (\gradv
\times \cv)$.  The latter term is present in $2D$ films whereas the former
is
not.  It is responsible for two-dimensionally modulated structures in
smectic-$C$ films \refs{\SL,\SEL}.  In three-dimensional
smectic-$C^*$ phases, the $\cv \cdot (\gradv \times \cv) \sim \nabla_z
\theta$
term is dominant, and the second term can often be ignored.  When, however,
the coefficient of $\nabla_z \theta$ is small or zero, the other term
becomes important and can lead to interesting two-dimensional structures
similar to those predicted for films and to slowly rotating versions of
these structures.  Indeed such structures have recently been
observed \ref\czech{E.~Gorecka {\sl et. al.}, Phys. Rev. Lett. {\bf 75}, 4047
(1995).}.

\newsec{Chiral Membranes and Lyotropic Phases}

\subsec{Membrane Model and Chiral Fluctuations}

Aliphatic molecules with polar heads and oily tails form membranes to
protect the oily tails from contact with water.
DMPC is a chiral molecule that can form a variety of lamellar and vesicular
structures.
In
the fluid, or $L_{\alpha}$, phase of these membranes, the molecular axes,
aligned along a unit director $\nv$, are parallel to the local membrane
normal
$\Nv$, and there is no macroscopic manifestation of chirality.  Indeed, if
$\nv$ is constrained to be parallel to $\Nv$, the phenomenological energy
describing a membrane with chiral molecules is identical to
that \ref\helfrich{W.~Helfrich, Z. Naturforsch. {\bf 28C}, 693 (1973);
P.~Canham, J. Theor. Biol. {\bf 26}, 61 (1970).} describing a membrane with
achiral molecules.  This
\nref\NelsonPowers{P.~Nelson and T.~Powers, Phys. Rev. Lett. {\bf 69} (1992)
3409;
J. Phys. II (Paris) {\bf 3} (1993) 1535.}
is because the curl of a vector normal to a surface is always zero.
When $\nv$ tilts relative to $\Nv$, as it does in the
$L_{\beta^{\prime}}$ phases, there is a component of $\nv$ in the
membrane's tangent plane, and there are chiral surface terms that have
been discussed extensively \refs{\SEL,\NelsonPowers}.  The
$L_{\alpha}$ phase of a membrane is similar to the smectic-$A$ phase of
a thermotropic liquid crystal: it expels twist, but it does not destroy
the underlying forces tending to produce twist. There are chiral
fluctuations in the $L_{\alpha}$ phase produced by fluctuations of the
director into the tangent plane.  To describe these fluctuations, we
treat the molecular director and membrane shape as independent
parameters.  Then the phenomenological free energy for a membrane can be
written as
\eqn\efreeagain{
F = F_h + F_n + F_{\rm Ch} + F_c ,
}
where
\eqn\efh{
F_h = {1\over 2} \kappa \int dS\, \left({1 \over R_1} + {1 \over
R_2}\right)^2 = {1\over 2} \kappa \int dS\, ( \gradv \cdot \Nv)^2
}
is the Helfrich-Canham energy \helfrich\ for an isolated membrane,
\eqn\efc{
F_c = - \gamma \int dS\, (\nv \cdot \Nv )^2
}
favors $\nv$ parallel to $\Nv$,
and $F_n$ and $F_{\rm Ch}$ are, respectively, the membrane versions of the
Frank director free energy and the chiral free energy:
\eqn\efn{\eqalign{
F_n = {1\over 2}\int dS\,&\bigg\{K_1 (\gradv_{\perp} \cdot \nv )^2 + K_2
(\nv\cdot\gradv_{\perp} \times \nv )^2\cr&\qquad
+K_3\left[\nv\times\left(\gradv_{\perp}\times
\nv\right)\right]^2\bigg\}\cr
}}
and
\eqn\efch{
F_{\rm Ch} =  K_2k_0\int dS\, \nv \cdot \left(\gradv_{\perp} \times \nv
\right) ,
}
where $\nabla_{\perp, i} = (\delta_{ij} - N_i N_j )\nabla_j$ is the
gradient restricted to the tangent plane.  Note that if $\gamma$ is
large locking $\nv$ parallel to $\Nv$,
then the effective Helfrich bending rigidity is $\kappa + K_1$.  In
the Monge gauge, we specify points on a membrane by their height $h(\xv)$
above a two-dimensional plane with co\"ordinates $\xv$. Then to linear order
in
$h$, we have $\Nv = ( - \gradv_{\perp} h , 1 )$, and
\eqn\efreexxx{
\eqalign{F&={1\over 2}\int d^2\!x\,
\bigg\{\kappa ( \nabla_{\perp}^2 h)^2 + K_1
(\gradv_{\perp} \cdot \delta \nv )^2 +K_2 (\nv\cdot\gradv_{\perp} \times
\delta \nv )^2\cr & \qquad\qquad\qquad+\left[\nv\times
\left(\gradv_{\perp}\times\nv\right)\right]^2\bigg\}\cr&
 \qquad + {1\over 2}\gamma \int d^2x (\gradv_{\perp} h + \delta \nv )^2
+K_2k_0 \int d^2 x ( \gradv_{\perp} \times \delta \nv)_z\cr .
}}
This free energy yields chiral fluctuations similar to those of the
smectic-$A$ free energy of \SmAfluc .

Since there are fluctuations in lamellae that contribute to the rotary
power, it is reasonable to believe that
an isotropic vesicular phase will have a fluctuation rotary power as well.
Since each vesicle will
undergo chiral fluctuations, the rotary power should, again, be
greater than the sum of microscopic contributions
from individual molecules.  It would be interesting to study the
rotary power of chiral amphiphiles as a function of their
concentration.  The lamellar $L_{\alpha}$ will have a chiral
fluctuations and a rotary power similar in form to those of a
thermotropic smectic-$A$ phase.  While it is typically believed that the there
is no in-plane order in lyotropic phases, the chiral fluctuations may still
lead to static chiral interactions which cause a tendency for cholesteric-like
twisting between neighboring ellipsoidal vesicles.

\subsec{Lyotropic Cholesteric Phase}

In the $L_{\alpha}$ phase,
membranes allow no average molecular twist if they form only closed
surfaces or
lamellae extending from one end of the sample to the other.
The only way to introduce twist into the membrane is to tear it:
it is possible to gain twist
energy
at the expense of edge energy in configurations that have the form of a
lamellar screw dislocation.  By adding a cosurfactant, such as
alcohol, the line energy will be reduced.  Then, as in the analysis
of all defect phases, there will be a trade-off between the energy of
destroying the integrity of the membrane and allowing chirality
into it.  As shown in Figure 2, the core of the dislocation is a tube filled
with water,
wrapped
helically by the edges of lamellar planes.  The molecular
director
exhibits a double twist configuration near the core that has negative
chiral
energy.

Using \efreexxx\ and the usual methods
to calculate the strain field of a dislocation, one can show
that the energy per turn of a membrane with a single screw dislocation with
pitch
$P$ and height profile $h(x,y) = (P/2 \pi) \tan^{-1}(y/x)$ is
\eqn\eenergy{
E = \gamma {P^2 \over 2 \pi} \ln (\lambda/\xi) - |K_2k_0| P + \tau P ,
}
where $\xi$ is the core radius, $\lambda = (K_2/ \gamma)^{1/2}$ is the
membrane
twist penetration depth, and $\tau$ is the line tension of the exposed edge.
Thus, if $|K_2k_0| - \tau$ is sufficiently large, it is energetically favorable
for
a membrane to create a screw dislocation with an exposed edge
rather than
to
maintain only flat or closed structures.

\bigskip
\centerline{\vbox{
\vbox{\hsize=4truein\baselineskip=0.16truein
\centerline{\psfig{figure=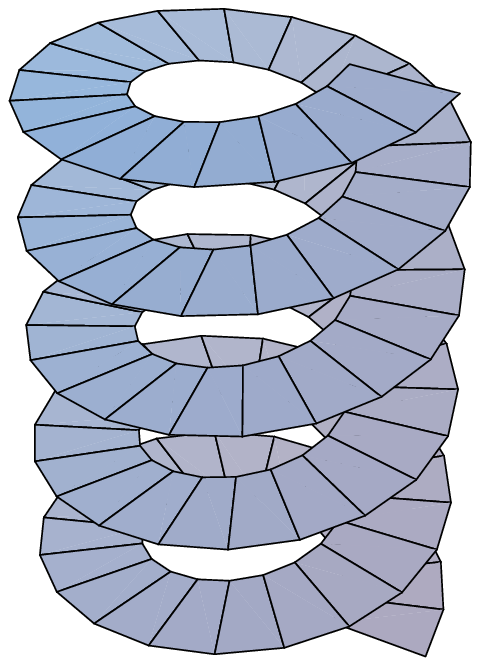}}
\noindent {\bf Figure 2.} This figure shows the structure of a chiral, screw
dislocation type defect in a lyotropic phase.  Note that the center
of the defect is filled with the solvent and there are exposed edges around
the center tube.
}}
}
\bigskip

In a lamellar phase, the pitch $P$ is identical to the layer spacing, which
is
set by a balance between a negative membrane-water surface tension and
either
Helfrich entropic or by electrostatic forces. Then, the energy per unit
length $\epsilon - E/P$ is identical to the energy per unit length of a
dislocation in a thermotropic smectic with $\tau$ playing the role of the
core energy.  As in thermotropic smectics, when $\epsilon$ becomes negative,
it
becomes favorable for screw dislocations to penetrate the lamellar phase
and
to create a lamellar TGB phase.  Local thermal fluctuations in dilute
lamellar phases are more violent than they are in thermotropic phases, and
one might expect the lamellar TGB phase to more susceptible to thermal
melting than its thermotropic cousin \ref\MKL{
I.~Bukharev, R.D.~Kamien, T.C.~Lubensky, and S.T.~Milner, {\sl in preparation}
(1996).}.
A melted TGB phase is a chiral line
liquid that is indistinguishable from a cholesteric phase \ref\LubKam{
R.D.~Kamien and T.C.~Lubensky, J. Phys. I (Paris) {\bf 3}, 2131 (1995).}.  We
can thus speculate that there should exist a lyotropic cholesteric phase
composed of a melted lattice of twist grain-boundaries  and possessing
local
lamellar ordering similar to that of the lamellar smectic-$A$ phase.

\newsec{Acknowledgments}
It is a pleasure to acknowledge stimulating discussions with R.~Meyer,
S.~Milner, P.~Nelson, T.~Powers, J.~Prost, D.~Roux,
C.~Safinya and especially D.~Walba, who brought these problems to
our attention.
This work was supported by NSF Grants DMR94-23114 and DMR91-22645.
\listrefs
\bye